\documentclass[a4paper,fleqn]{cas-sc}
\usepackage{gensymb}
\usepackage[numbers]{natbib}
\def\tsc#1{\csdef{#1}{\textsc{\lowercase{#1}}\xspace}}
\tsc{WGM}
\tsc{QE}
\tsc{EP}
\tsc{PMS}
\tsc{BEC}
\tsc{DE}
\DeclareUnicodeCharacter{2212}{-}

\begin{document}
\let\WriteBookmarks\relax
\def\floatpagepagefraction{1}
\def\textpagefraction{.001}
\shorttitle{}
\shortauthors{S. Sharma et~al.}
%\begin{frontmatter}

% \title [mode = title]{Investigation of jet noise using one-dimensional turbulence}   
%%MK%%\title [mode = title]{On the reduced-order framework based on one-dimensional turbulence for jet noise}   
\title [mode = title]{On a lower-order framework for jet noise prediction based on one-dimensional turbulence}   
% \tnotemark[1,2]

% \tnotetext[1]{This document is the results of the research
%   project funded by the National Science Foundation.}

% \tnotetext[2]{The second title footnote which is a longer text matter
%   to fill through the whole text width and overflow into
%   another line in the footnotes area of the first page.}

\author[1,2]{Sparsh Sharma}[type=editor,
                        auid=000,bioid=1,
                        prefix=,
                        role=,
                        orcid=0000-0002-6978-8663]
% \cormark[1]
\ead{sparsh.sharma@b-tu.de}
% \ead[url]{www.cvr.cc, cvr@sayahna.org}

\credit{Conceptualization, Methodology, Data curation, Writing - Original draft preparation}
\address[1]{Technische Akustik, Brandenburgische Technische Universität Cottbus-Senftenberg, Cottbus, Germany}

\author[2]{Marten Klein}
\credit{Conceptualization of this study, Methodology, Writing - review \& editing
}
\author[2]{Heiko Schmidt}[%
   role=,
   suffix=,
   ]
% \ead{cvr3@sayahna.org}
% \ead[URL]{www.sayahna.org}

\credit{Funding acquisition, Writing - review \& editing}
% Data curation, Writing - Original draft preparation
\address[2]{Lehrstuhl Numerische Strömungs- und Gasdynamik, Brandenburgische Technische Universität Cottbus-Senftenberg, Cottbus, Germany}

\author%
[3]
{Ennes Sarradj}
% \cormark[2]
% \fnmark[1,3]
% \ead{rishi@stmdocs.in}
% \ead[URL]{www.stmdocs.in}
\credit{Writing - review \& editing}
\address[3]{Technische Akustik, Technische Universität Berlin, Berlin, Germany}

\cortext[cor1]{Corresponding author}
% \cortext[cor2]{Principal corresponding author}
% \fntext[fn1]{This is the first author footnote. but is common to third
% \fntext[fn2]{Another author footnote, this is a very long footnote and
%   it should be a really long footnote. But this footnote is not yet
%   sufficiently long enough to make two lines of footnote text.}

% \nonumnote{This note has no numbers. In this work we demonstrate $a_b$
%   the formation Y\_1 of a new type of polariton on the interface
%   between a cuprous oxide slab and a polystyrene micro-sphere placed

\begin{abstract}
Noise prediction requires the resolution of relevant acoustic sources on all scales of a turbulent flow.
High-resolution direct numerical and large-eddy simulation would be ideal but both are usually too costly despite developments in high-performance computing.
Lower-order modeling approaches are therefore of general interest. 
A crucial but standing problem for accurate predictive modeling is the estimation of missing noise from the modeled scales.
In this paper we address this problem by presenting a novel lower-order framework that couples the one-dimensional turbulence model to the Ffowcs-Williams and Hawkings approach for prediction of the far-field noise of a subsonic turbulent round jet.  
\end{abstract}

% \includegraphics{figs/grabs.pdf}

% \begin{highlights}
% \item Formulation of a lower-order framework for high-Reynolds-number turbulent noise prediction based on the stochastic one-dimensional turbulence (ODT) model and the Ffowcs-Williams and Hawkings (FW-H) equation.
% \item Application of the stand-alone lower-order framework to turbulent subsonic round jets. 
% %%%MK: CHECK terminology: "angle \theta" = "view angle"?
% %%checked: it's 'radiation angle'
% \item Accurate prediction of noise similarity spectra at large radiation angles due to fine-scale turbulence in a subsonic turbulent round jet.
% %%%MK: Replaced by "Application ..." above. My concern: Can you live up to this promise in the paper? The CPU-h etc. are not given anywhere ... 
% %\item The proposed approach based on ODT leads to major cost reduction compared to DNS / LES through lower spatial dimensionality, thereby, making higher Reynolds number simulations possible.
% \end{highlights}


\maketitle

\section{Introduction}
Numerical evaluation of the performance of new noise-reducing concepts relies on robust modelling methods for the sources of the participating acoustic processes.
Direct numerical simulation (DNS) achieves this but its resolution requirements limit its applicability to mildly turbulent flows. 
Large-eddy simulation (LES) coupled with acoustic analogies has become the method-of-choice for jet noise prediction \citep{Lyrintzis2020}. 
In a typical LES, about 80\% of the turbulent kinetic energy are resolved and 20\% are modelled by a subgrid-scale (SGS) averaging approach due to which fine-scale information has been filtered out.
Fine-scale turbulence, however, is considered as the dominant noise source component of turbulent mixing noise in subsonic jets because of the absence of Mach wave radiation \citep{Tam2008}. \citet{Bres2019} in their recent review of jet noise mention that the estimation of the missing noise from the unresolved scales is one of the remaining challenges in modelling of jet noise. 

\noindent
We address this challenge by formulating a lower-order framework that aims to resolve all relevant scales of a turbulent flow with the aid of the one-dimensional turbulence (ODT) model \citep{kerstein1999}.
ODT achieves cost reduction by resolving the flow on a quasi-one-dimensional domain along which turbulent advection is modeled by a stochastic process but molecular diffusion is directly resolved.
In comparison to LES or Reynolds-averaged Navier--Stokes (RANS) modelling, there is no need for closure and, thus, no eddy viscosity is involved. 

\noindent
The framework is validated as follows. The near-field acoustic sources of a turbulent subsonic unheated round jet are resolved with ODT as a stand-alone tool.
After that, a ODT surrogate pressure is defined with which we estimate the far-field noise using the Ffowcs-Williams and Hawkings (FW-H) approach. The novelty is the coupling of established methods by the surrogate pressure by which effects of fine-scale turbulence on jet noise can be investigated. 

\noindent 
The paper is organised as follows. In Section \ref{sec:1}, a theoretical background of ODT is presented. In Section \ref{sec:3}, the FW-H approach along with the near-field and far-field results is discussed. Conclusions are drawn in Section \ref{sec:4}.

\section{Formulation of the lower-order framework and case set-up}
\label{sec:1}
\subsection{Overview of the ODT model}

There are two alternative ODT formulations, T-ODT and S-ODT \cite{kerstein1999}. T-ODT is for temporally (like channel flow) and S-ODT for spatially developing flows (like jets). For the study of noise generated by a turbulent round jet, we utilise S-ODT in its recent extension to flows with axially symmetric flow statistics \cite{Lignell2018} as sketched in Fig.~\ref{fig:f11a}. For the simplest imaginable case of a low-Mach number flow with almost constant fluid density and viscosity the ODT streamwise momentum equation for a round jet may be written as
\begin{equation}
u \dfrac{\partial \boldsymbol{\rm u}}{\partial x} + v \dfrac{\partial \boldsymbol{\rm u}}{\partial r} + EE
 = \dfrac{1}{r} \dfrac{\partial}{\partial r}\left( \nu r \dfrac{\partial \boldsymbol{\rm u}}{\partial r} \right),
 \label{eq:mom}
\end{equation}
where $\boldsymbol{\rm u}=(u,v,w)^\mathrm{T}$ is the velocity vector with axial ($u$), radial ($v$) and azimuthal ($w$) components, $x$ is the axial and $r$ the radial coordinate, $\nu$ is the kinematic fluid viscosity, and $EE$ represents the stochastic ODT eddy events.
The numerical solution procedure involves a fully-adaptive finite-volume discretisation on a Lagrangian grid with piecewise parabolic advancement of deterministic processes. For implementation details, we refer the reader to \citep{Lignell2018}.

\noindent
Eddy events are a corner stone of ODT and defined by the random variables radial eddy size and location for a given flow state \citep{kerstein1999}.
Sampling is performed by a thinning-and-rejection algorithm in which the acceptance probability is governed by the momentary flow on the radial ODT domain that is located at some downstream position~$x$.
An S-ODT simulation provides a synthetic but statistically representative quasi-two-dimensional `snapshot' of the flow.
An ensemble of these realisations (`snapshots') is analogous to an ensemble of instantaneous physical flow states with regard to the statistics that can be gathered \cite{Lignell2018}.

%%%%%%%%%%%%%%%% TO BE REMOVED
%fix it sparsh

% In the ODT model, a stochastic implementation of 1-D eddy events to model the effect of nonlinear advection and fluctuating pressure gradient terms is coupled to the deterministic solution of a 1-D diffusion evolution equation. The effects of turbulent transport due to eddies on the 1-D property profiles of the flow are modelled by instantaneous eddy events and the deterministic diffusion process occurs between these events. The governing equations in ODT can be expressed in a temporal formulation (T-ODT) as well as a spatial formulation (S-ODT). The latter one is used for the present investigation and can be expressed as

% \begin{equation}
%     u\frac{\partial u}{\partial x}+v\frac{\partial u}{\partial y}+EE=\nu \frac{\partial^2 u}{\partial y^2},
%     \label{eq:eq2}
% \end{equation}

% \noindent
% where the first term represents the local changes of the velocity, $u$ with respect to the longitudinal direction $x$, $v$ the vertical advecting velocity, $EE$ the stochastic eddy events term, and $\nu$ the kinematic viscosity. These eddy events occur through the instantaneous displacement of the fluid elements to represent a turbulent stirring motion and are implemented by using the triplet map. For the complete model formulation, refer to \cite{kerstein1999}. In S-ODT, despite having the nominal steadiness of each flow realisation, an ensemble of such realisations is analogous to an ensemble of instantaneous physical flow states with regard to the statistics that can be gathered \cite{Lignell2018}.
%%%%%%%%%%%%%%%% 

\subsection{A surrogate pressure for mixing noise estimation}
\begin{figure}[t]
\centering
        \begin{subfigure}[b]{0.46\textwidth}
                \centering
                \includegraphics[width=\linewidth]{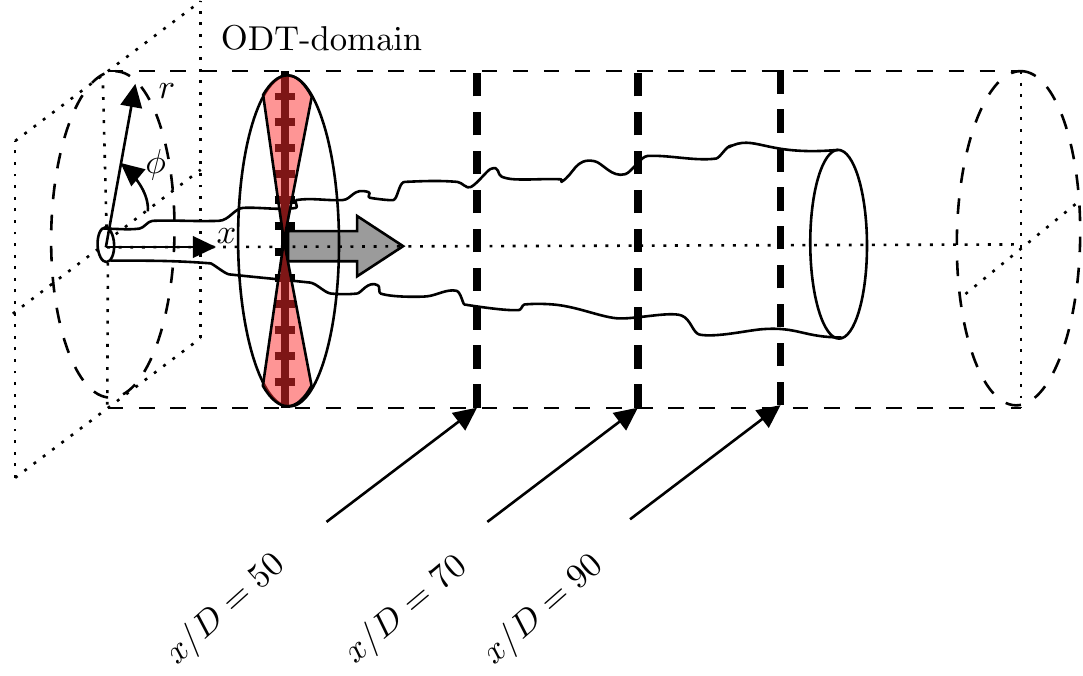}
                \caption{}
                \label{fig:f11a}
        \end{subfigure}\hfill
        \begin{subfigure}[b]{0.46\textwidth}
                \centering
                \includegraphics[width=\linewidth]{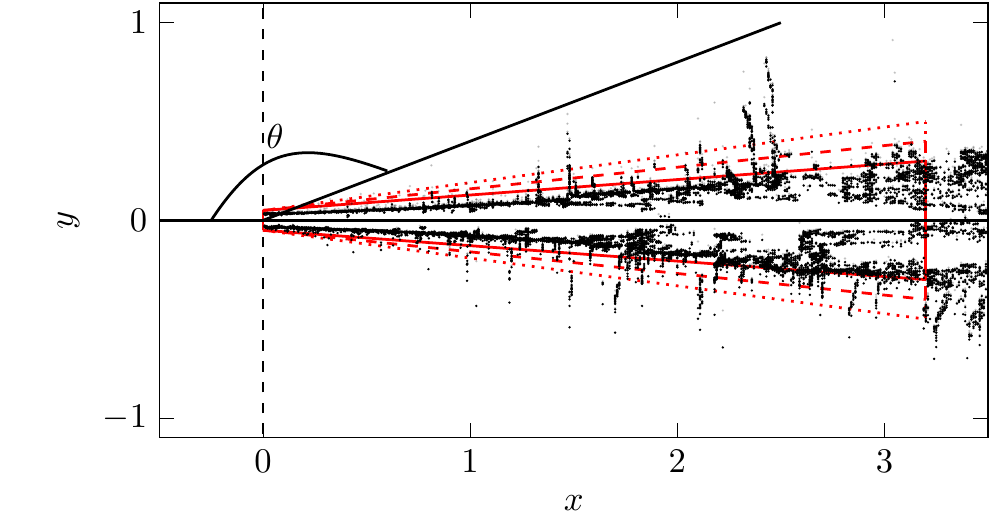}
                \caption{}
                \label{fig:f11b}
        \end{subfigure}
        \caption{(a) Schematic of the round jet (not to scale). Coordinates ($r,\phi,x$) denote the radial, azimuthal, and axial directions.
        %ODT simulations are conducted in the central ($x,y$)-plane with a moving ODT line, 
        (b) Acoustics data surface (red) for the FW-H method superimposed to a sequence of eddy events from a single S-ODT realisation for a length of $50D$.}\label{fig:f11}
\end{figure}

The ODT model resolves all relevant scales but does not solve for the (kinematic) pressure, which is one of the reasons why ODT is relatively inexpensive \citep{kerstein1999}. 
The model, instead, employs a conservative mapping and a kernel mechanism in order to allow energy redistribution among the components of the velocity vector. %%%MK%%% \cite{Kerstein2001}.
These operations aim to model the effects of turbulent advection and fluctuating pressure forces, respectively, which may be written as
\begin{equation}
EE:\quad \boldsymbol{\rm u}(r) \to \boldsymbol{\rm u}\big(f(r)\big) + \boldsymbol{\rm c}\, K(r).
\label{eq:eddy}
\end{equation}
In the above equation, $f(r)$ denotes the triplet map, $K(r)$ a mapping kernel, and $\boldsymbol{\rm c}=(c_1,c_2,c_3)^\mathrm{T}$ [s$^{-1}$] the vector of kernel coefficients.
Both $f(r)$ and $K(r)$ are prescribed mapping functions with dimension [m]. Information about kinematic pressure and, hence, the sources for mixing noise, are contained in the $c_i$. 
(See \citep{Lignell2018} for computation of $c_i$ in cylindrical coordinates.)
%%
%%\noindent
The coefficients $c_i$ are localised, that is, effective only for the volume fraction of the ODT domain that is occupied by an eddy event.
In order to estimate mixing noise, we collect the sources in terms of $c_i$ for all eddy events that cross a predefined control surface as shown in Fig.~\ref{fig:f11b}. 
Eddy events have a given radial but no axial extend.
We therefore attribute a streamwise extend, $l_e$, by a time-to-space transformation, $l_e = \tau_e\,u_e$, of the eddy turnover time, $\tau_e$, that is available from the sampling procedure \citep{kerstein1999,Lignell2018}, and the local streamwise velocity, $u_e=u(r_e)$.
An effective kinematic surrogate pressure, $\pi$ [m$^2$/s$^2$], is estimated with dimensional arguments from $c_i$ and $l_e$ as
\begin{equation}
    \pi = p/\rho = |\boldsymbol{\rm c}|^2\,l_e^2.
\end{equation} 

\noindent
An ensemble of independent ODT realisations is used to obtain a synthetic but statistically representative time history of the fluctuating surrogate pressure on the predefined control surface. This time-dependent surrogate pressure is then used to predict the acoustic pressure using the FW-H approach \cite{Williams1969}.

% \begin{figure}[!t]
%     \centering
%     \includegraphics[scale=0.65]{figs/f6.pdf}
%     \caption{Schematic of the round jet (not to scale). The coordinates ($x,y,z$) denote the streamwise, wall-normal and spanwise directions. ODT simulations are conducted in the central ($x,y$)-plane with a moving ODT line.}
%     \label{fig:f1}
% \end{figure}

\subsection{Unheated round jet set-up}
We consider a jet of air that issues into ambient airthrough a $D=0.062\,\mathrm{m}$ diameter round nozzle as sketched in Fig.~\ref{fig:f11a}. The simulations are conducted at low Mach numbers, $M=0.21$, $0.24$, $0.26$, $0.29$, and $0.31$. 
ODT simulations were performed for these Mach numbers %jet exit velocities %with varying ODT parameters as listed in Table. \ref{tab1}. ODT simulations were performed. 
by keeping the ODT model parameters fixed at $C=5.25$ and $Z=400$ \cite{Lignell2018}.
A parameter sensitivity analysis performed for $M=0.31$ revealed that turbulent dispersion of the jet increases with the turbulence intensity parameter, $C$, while the viscous cut-off parameter, $Z$, has no notable influence on the jet. 

\noindent
The initial velocity profile used in the ODT simulations is a smoothed top-hat profile \cite{Lignell2018},
\begin{equation}
    u_0(r)=u_{\rm min}+\Delta u \frac{1}{2}\left ( 1+\tanh\left ( \frac{2}{\delta}\left ( r-r_{\rm c1} \right ) \right ) \right )\frac{1}{2}\left ( 1+\tanh\left ( \frac{2}{\delta}\left ( r_{\rm c2}-r \right ) \right ) \right ),
    \label{eq:eq1}
\end{equation}
in which a $\tanh$ function of width $\delta = 0.1D$ is used on either side of the jet to avoid discontinuities and $r_{c1/c2}=\pm D/2$ denote the inflection points of the $\tanh$ transition. 
A small co-flow velocity, $0<u_\mathrm{min}\ll\Delta u$, is necessary to assure existence of an S-ODT solution.
We note that \cite{Albertson1950} used a Gaussian inflow profile, $u_0(r)\sim \exp(r^2)$. %, whereas we use the $\tanh$ profile, Eq.~\eqref{eq:eq1}, as in the validated ODT model set-up described in \cite{Lignell2018}.
Details related to the inflow profile used affect only a small region, $0<x\lesssim O(D)$, behind the nozzle where the jet is dominated by flow instabilities. This region, however, has negligible contribution to the mixing noise radiated. 

% \begin{table}[width=.9\linewidth,cols=5,pos=b]
% \caption{Test cases based on different Mach numbers and ODT model parameters.}\label{tbl1}
% \begin{tabular*}{\tblwidth}{@{} LLLLL@{} }
% \toprule
% M & D & Re & C & Z\\
% \midrule
% 0.21 & 0.06223 & 2.4$\times 10^5$ & 400  & 5.25 \\
% 0.24 & 0.06223 & 2.8$\times 10^5$ & 400  & 5.25 \\
% 0.26 & 0.06223 & 3.0$\times 10^5$ & 400  & 5.25 \\
% 0.29 & 0.06223 & 3.4$\times 10^5$ & 400  & 5.25 \\
% \multirow{ 3}{*}{0.31} & \multirow{ 3}{*}{0.06223} & \multirow{ 3}{*}{3.6$\times 10^5$} & 200 & 5.25 \\
% & & & \multirow{ 3}{*}{400} & 5.25 \\
% & & & & 9.0 \\
% & & & & 12.0 \\
% & & & 600 & 5.25 \\
% \bottomrule
% \end{tabular*}
% \label{tab1}
% \end{table}

\section{Results}
\label{sec:3}
\subsection{Near-field ODT simulations}
\begin{figure}[t]
\centering
        % \begin{subfigure}[b]{0.32\textwidth}
        %         \centering
        %         \includegraphics[width=\linewidth]{figs/f1a.pdf}
        %         \caption{}
        %         \label{fig:f2a}
        % \end{subfigure}\hfill
        \begin{subfigure}[b]{0.32\textwidth}
                \centering
                \includegraphics[width=\linewidth]{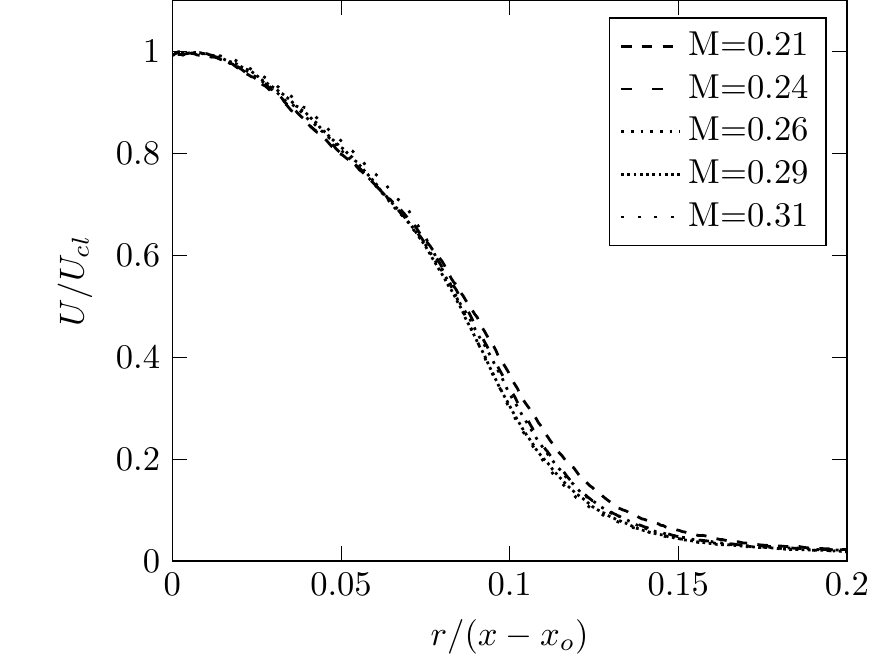}
                \caption{}
                \label{fig:f2b}
        \end{subfigure}%\hfill
        \begin{subfigure}[b]{0.32\textwidth}
                \centering
                \includegraphics[width=\linewidth]{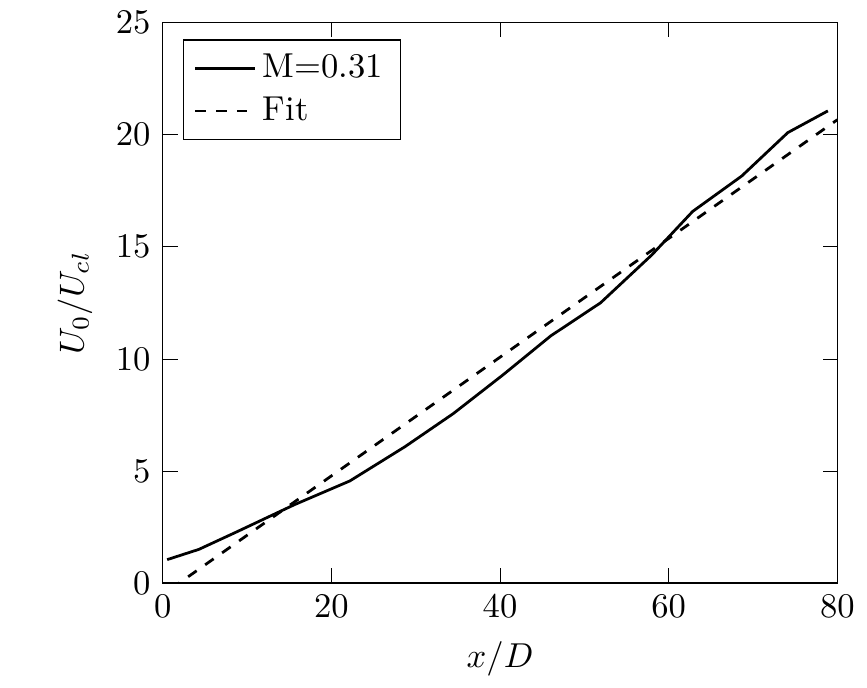}
                \caption{}
                \label{fig:f2c}
        \end{subfigure}
        \caption{(a) Normalised radial profiles of the mean axial velocity at $x/D=50$ for all Mach numbers investigated using fixed model parameters $C=5.25$ and $Z=400$. (b) Normalized mean centreline axial velocity versus downstream location for $M=0.31$.}\label{fig:f2}
\end{figure}

A total of 5000 independent ODT realizations were performed for each case and results were ensemble-averaged. %All quantities are normalized consistent with jet similarity scaling. 
Downstream locations are normalized by the jet diameter $D$, and radial locations are normalized by $(x-x_0)$ where $x$ is the downstream location and $x_0=4D$ is the virtual origin \cite{Lignell2018}. In the figure, $U_0$ is the mean jet exit velocity and $U_{cl}$ is the local mean axial velocity at the centreline. 
%Fig.~\ref{fig:f2a} shows radial profiles of the mean axial velocity for all jet exit velocities investigated at $x=50D$. The mean axial velocity normalized by the local centerline value is shown in Fig.~\ref{fig:f2b}.
Fig.~\ref{fig:f2b} shows radial profiles of the normalized mean axial velocity at $x=50D$ for all jet exit velocities investigated.
ODT correctly predicts a self-similar mean state.
In Fig.~\ref{fig:f2c}, the similarity scaling gives a nominally linear profile where $U$ decays like $1/x$ (dashed line in Fig.~\ref{fig:f2c}). %%%MK: Why increase in Fig.2c? Please add the 1/x trend (dashed) in Fig.2c.
These ODT simulations are consistent with the validation in \cite{Lignell2018} and form the basis for noise prediction.

\subsection{Far-field noise predictions}
\begin{figure}[!t]
        \centering
        \begin{subfigure}[b]{0.44\textwidth}
            \centering
            \includegraphics[width=\textwidth]{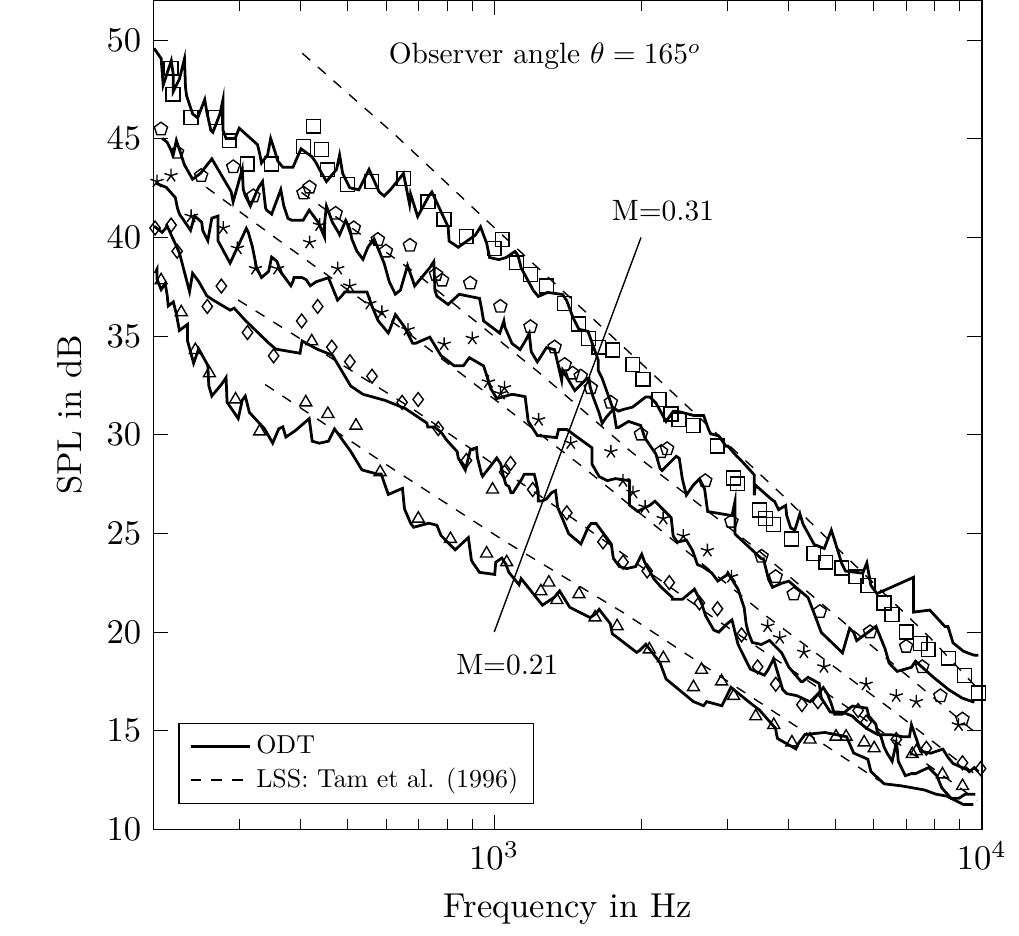}
            \caption[]%
            {} 
            \label{fig:fLa}
        \end{subfigure}
        \hfill
        \begin{subfigure}[b]{0.44\textwidth}  
            \centering 
            \includegraphics[width=\textwidth]{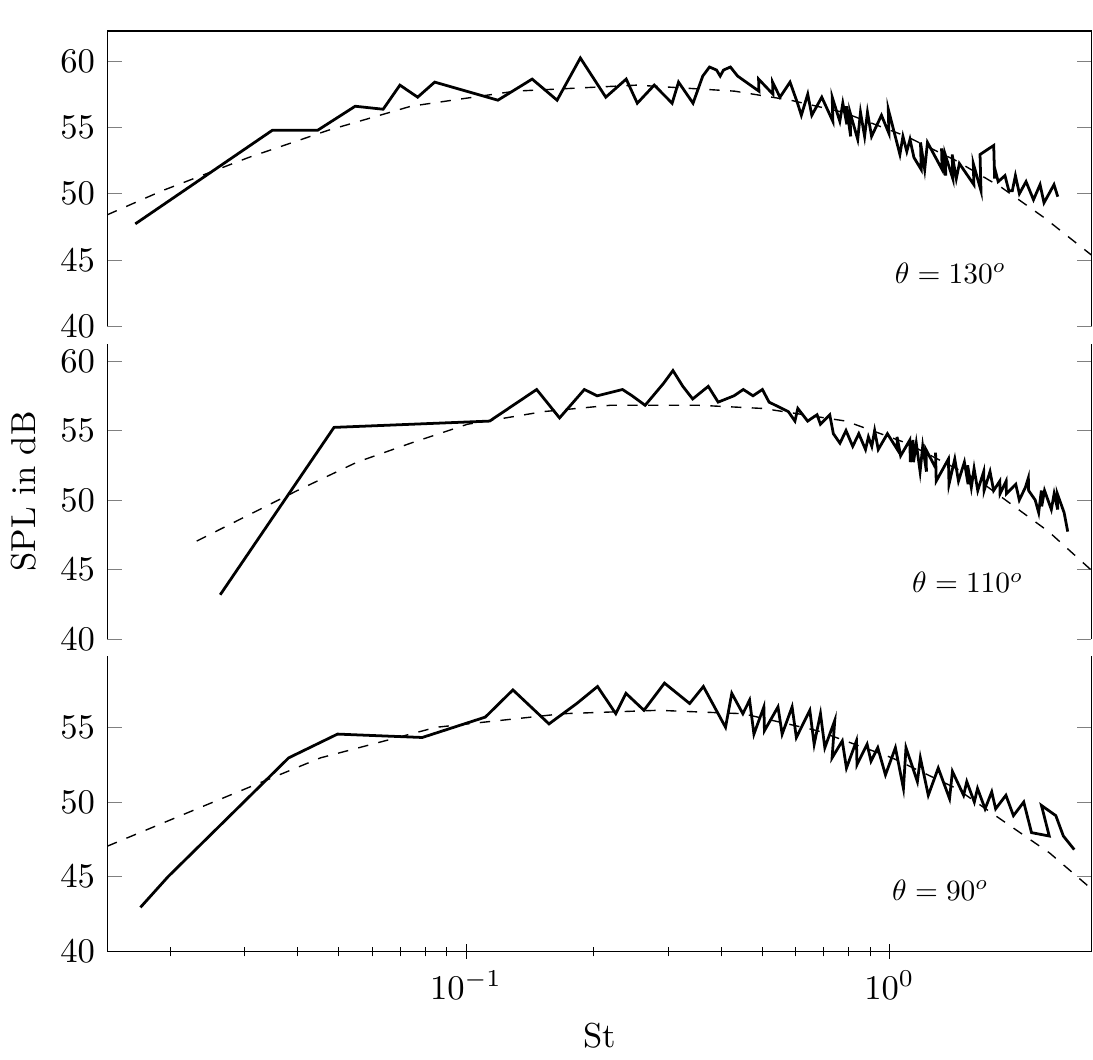}
            \caption[]%
            {}
            \label{fig:fLb}
        \end{subfigure}
        \caption{(a) Spectra of the sound-pressure level (SPL) of round jets at $165\degree$ for various Mach numbers. Measured data (symbols) are from \cite{Viswanathan2007}. (b) Comparison of predicted (solid) with the fine-scale similarity (dashed) spectra for $M=0.31$.}
        \label{fig:fL}
\end{figure}

% s and results were recorded representing the time history of pressure estimator term on the predefined control surface around the jet. 
The far-field noise prediction is facilitated by the frequency-domain permeable formulation of the FW-H equation \citep{Williams1969}, which is a surface integral method that uses the near field information gathered over a surface enclosing as many as possible of the noise sources. The quadrupole contributions from the volume-distributed noise sources outside of the FW-H surface are neglected. The flow field and surrogate pressure on the control surface, shown in Fig.~\ref{fig:f11b}, is gathered in every ODT realisation. 
%%Use this when you include the FW-H equation%%
%The source terms are calculated and transformed into the frequency-domain using the Fourier transform. The integrals are then evaluated for each point of an acoustic data surface. Lastly, an inverse Fourier transform is used to recover the acoustic signal in the time domain.
%In a post-processing step, the far-field radiated sound is calculated using the FW–H surface data via analytically known Green’s function for a stationary or uniformly moving ambient medium. The accurate implementation of the FW-H method is based on the correct definition of the FW-H surface enclosing the noise sources. A prime example of this anomaly is the treatment of the downstream end of the FW-H surface. 
The far-field noise is calculated from the nozzle exit for three different FW-H surfaces consisting of a conical surface extending to $x/D = 50$ with different apex angles.
The corresponding slopes as seen in the $(x,r)$-plane are $0.11$, $0.14$, and $0.17$. Here, these slopes are chosen based on estimates of the jet spreading rate \citep{Albertson1950} that ODT correctly captures. %\citep{,Lee2003,Bres2020}. 
Observers are located at a constant sideline at a distance of $R=4.572\,\mathrm{m}$ from the jet axis and at radiation angle $\theta=165\degree$ at a distance of $R=2.79\,\mathrm{m}$ \citep{Viswanathan2007}. All angles are measured from the upstream-directed jet inlet axis. Fig.~\ref{fig:fLa} shows the narrow-band spectra at radiation angle $\theta=165\degree$ for all Mach numbers investigated. Fig.~\ref{fig:fLb} shows the noise spectra of for $M=0.31$ for three radiation angles.

% The placement of FW-H surface, and the spatial and temporal resolution of the data saved on FW-H surface are important. If the FW-H surface lies inside the turbulent flow, the data collected on it do not include the sound produced by the turbulent flow lying outside of it and (vigorous) crossing of turbulence across the surface is a contributor to spurious acoustics. On the other hand, if the FW-H surface is placed too far away from turbulent jet, predictions based on it might also be incorrect because the acoustic data recorded on that distant surface might be corrupted by significant numerical errors. 

% \begin{figure}[!t]
%     \centering
%     \includegraphics[scale=0.75]{figs/f4.pdf}
%     \caption{Acoustics data surface (in red) for FW-H method superimposed over a single S-ODT realisation for a length of $50D$.}
%     \label{fig:f4}
% \end{figure}

\section{Conclusion}
\label{sec:4}
We address missing noise estimation by presenting the formulation and first results of a lower-order numerical framework.
This framework is based on the stochastic ODT model for the turbulent near-field that is coupled by a surrogate pressure to the FW-H equation for the acoustic far-field.
We have applied the model to turbulent round jets that are dominated by mixing noise and obtained robust, reasonable, and feasible noise predictions for the validated near-field model.
For large radiation angles, the predicted noise (Fig.~\ref{fig:fLa}) agrees well with large-scale similarity spectra \cite{Tarn1996} and exhibits the expected monotonic increase in the sound-pressure level \cite{Viswanathan2007}.
For small radiation angles, the model prediction (Fig.~\ref{fig:fLb}) matches reference fine-scale similarity spectra \cite{Tarn1996}.  %%%MK: FSS ref.
The results obtained suggest that small-scale resolution is crucial for closing the missing-noise gap..

\printcredits

\bibliographystyle{model1-num-names}
% \bibliographystyle{cas-model2-names}

\bibliography{cas-refs}
\end{document}